\documentclass[journal]{IEEEtran}

\usepackage{epstopdf}

\usepackage{graphicx}
\usepackage{amsmath}
\usepackage{amsfonts}
\usepackage{amssymb, color}

\newtheorem{theorem}{Theorem}

\begin{document}

\title{Wireless Networks with Energy Harvesting and Power Transfer: Joint Power and Time Allocation }
\author{Zoran Hadzi-Velkov, Ivana Nikoloska, George K. Karagiannidis, and Trung Q. Duong
\thanks{Z. Hadzi-Velkov and I. Nikoloska are with the Faculty of Electrical Engineering and Information Technologies, Ss. Cyril and Methodius University, 1000 Skopje, Macedonia (email: \{zoranhv, ivanan\}@feit.ukim.edu.mk).}
\thanks{G. K. Karagiannidis is with the Department of Electrical and Computer Engineering, Aristotle University of Thessaloniki, 54 124 Thessaloniki, Greece, (e-mail: geokarag@auth.gr).}
\thanks{T. Q. Duong is with the School of Electronics, Electrical Engineering and Computer Science, Queen's University Belfast, Belfast BT7 1NN, U.K. (e-mail: trung.q.duong@qub.ac.uk).}
}

\markboth{Accepted for journal publication}{Shell \MakeLowercase{\textit{et al.}}: Bare Demo of IEEEtran.cls for Journals} \maketitle

\vspace{-5mm}
\begin{abstract}
In this paper, we consider wireless powered communication networks which could operate perpetually, as the base station (BS) broadcasts energy to the multiple energy harvesting (EH) information transmitters. These employ "harvest then transmit" mechanism, as they spend all of their energy harvested during the previous BS energy broadcast to transmit the information towards the BS. Assuming time division multiple access (TDMA), we propose a novel transmission scheme for jointly optimal allocation of the BS broadcasting power and time sharing among the wireless nodes, which maximizes the overall network throughput, under the constraint of average transmit power and maximum transmit power at the BS. The proposed scheme significantly outperforms "state of the art" schemes that employ only the optimal time allocation. If a single EH transmitter is considered, we generalize the optimal solutions for the case of fixed circuit power consumption, which refers to a much more practical scenario.
\end{abstract}


\vspace{-5mm}

\section{Introduction}
Energy harvesting (EH) is considered a revolutionary technology for energy-constrained wireless networks, such as sensor and ad-hoc, due to its capability to provide everlasting power supply [\ref{lit1}]-[\ref{lit2}]. Performance of these systems is maximized by adapting the output powers of the EH transmitters given the energy causality constraint [\ref{lit3}]-[\ref{lit4}]. However, EH from the environment (e.g., solar or wind) is an intermittent and uncontrollable process. In order to maintain reliable EH-based communication, dedicated far-field radio frequency (RF) radiation is used as energy supply for EH transmitters, which is known as wireless power transfer (WPT) [\ref{lit6}]. The WPT can be realized as a simultaneous wireless information and power transfer [\ref{lit7}], or, alternatively, over a dedicated (either time or frequency) channel for energy transfer. The latter option gives rise to the so-called wireless powered communications networks (WPCNs), which typically transmit using the time-division multiple access (TDMA) [\ref{lit8}]-[\ref{lit12}]. An example WPCN may be a sensor network consisted of a base station (BS) and multiple EH sensors deployed in a hostile environment, such that they send information (e.g., telemetry) over the uplink and receive energy broadcasted by the BS over the downlink.

In [\ref{lit8}], the authors determine the optimal TDMA scheme among the half-duplex nodes (either BS or EHNs), depending on the channel fading states. The optimal time-sharing in WPCNs with separated frequency channels for energy broadcast and information transmissions (IT), were studied in [\ref{lit9}]. Furthermore, [\ref{lit10}] studies the WPCN with full-duplex nodes, where the BS is equipped with two antennas. The authors in [\ref{lit11}] consider WPCNs, where the nodes choose between two power levels, i.e., a constant desired power, or a lower power when its EH battery has stored insufficient energy. Finally, in [\ref{lit12}] a three-node relaying system was considered, where both source and relay harvest energy from the BS, by using WPT.

The above-referenced works analyzed WPCNs (except for [\ref{lit11}]) and proposed optimal (either time or power) allocation schemes that aim at maximizing the achievable sum information rate (i.e., throughput in WPCNs). In this paper, we consider a similar WPCN model as in [\ref{lit8}] and propose a novel optimal scheme, which jointly allocates the BS broadcasting power and the time sharing among the wireless nodes. Our proposed scheme significantly improves the overall WPCN throughput, compared to [\ref{lit8}], which assumes a constant BS transmit power. For the point-to-point scenario, the proposed scheme is generalized to account for the fixed (circuit) power consumption by the EHNs, which, to the best of authors' knowledge, has not been considered in the literature so far.

\vspace{-4mm}
\section{System and channel model}
As depicted in Fig. 1(a), we consider a WPCN in a fading environment, consisting of a half-duplex BS and $K$ half-duplex EHNs. The BS broadcasts RF energy to the EHNs, whereas the EHNs transmit information back to the BS. The EHNs are equipped with rechargeable EH batteries that harvest the RF energy broadcasted from the BS. The IT from EHNs to BS (IT phase), and the WPT from BS to EHNs (EH phase) are realized as successive signal transmissions using TDMA over a common channel, where each TDMA frame/epoch is of duration $T$. As depicted in Fig. 1(b), each (TDMA) epoch consists of an EH phase and an IT phase, whereas the IT phase itself consists of $K$ successive ITs from EHNs to the BS. We assume perfect synchronization among all the nodes.

The fading between the the BS and EHN $j$ ($1 \leq j \leq K$) is a stationary and ergodic random process, and follows the quasi-static block fading model, i.e., the channel is constant in each slot but changes from one slot to the next. The duration of one fading block is assumed equal to $T$, and one block coincides with a single epoch. For convenience, the downlink (BS-$\text{EHN}_j$) and the uplink ($\text{EHN}_j$-BS) channels are assumed reciprocal, and their fading power gains in epoch $i$ are denoted by $x_{ji}'$. They are normalized by the additive white gaussian noise (AWGN) power, yielding $x_{ji} = x_{ji}'/N_0$ with an average value of $\Omega_j = E[x_{ji}']/N_0$, where $E[\cdot]$ denotes expectation.

\begin{figure}[tbp]
\centering
\includegraphics[width=3.7in]{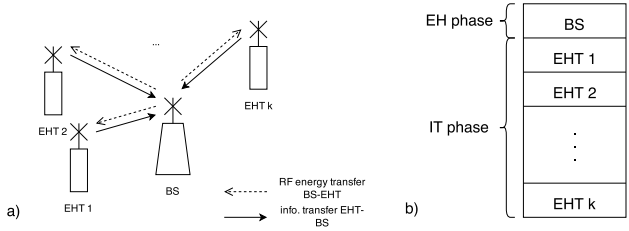}
\caption{(a) Wireless powered network with $K$ EHNs and a common BS, (b) TDMA epoch/frame structure} \vspace{-5mm}
\label{fig1}
\end{figure}

\vspace{-2mm}
\section{WPCN throughput maximization}
Let us consider $M \to \infty$ epochs. In epoch $i$, the BS transmits with power $p_i$. The duration of the EH phase is, $\tau_{0i} \, T$, and the IT duration of the EHN $j$ is $\tau_{ji} \, T$ ($1 \leq j \leq K$). Note that, $0 < \tau_{ji} < 1$ and $\sum_{j=0}^{K} \tau_{ji} = 1, \forall i$, where $\tau_{ji}, \forall ij$ are referred to as the time-sharing parameters.

We propose a scheme for jointly optimal allocation of $p_i$ and $\tau_{ji}$ so as to maximize the WPCN overall average throughput, subject to the average available BS transmit power, $P_{avg}$, and maximum BS transmit power, $P_{max}$. The proposed scheme is feasible if, in each epoch $i$, the BS has perfect knowledge of the instantaneous fading gains in all BS-$\text{EHN}_j$ channels.

During the EH phase of epoch $i$, the amount of harvested power by EHN $j$ is $E_{ji} = \eta_j x_{ji}  N_0 p_i \, \tau_{0i} \, T$, where $\eta_j$ is the energy harvesting efficiency of EHN $j$. Note, it is safe to assume that $\eta_j = 1$, because $\eta_j$ always appears together with $x_{ji}$ through the product $\eta_j x_{ji}$, which only affects the parameter $\Omega_j$, but the form of the end results is unaffected. During the IT phase, EHN $j$ completely spends its harvested energy, $E_{ji}$, for transmitting a complex-valued Gaussian codeword of duration $\tau_{ji} \, T$, comprised of $n \to \infty$ symbols, with an output power
\begin{equation}
P_j(i) = \frac{E_{ji}}{\tau_{ji} \, T} = \frac{N_0 p_i \, x_{ji} \, \tau_{0i}}{\tau_{ji}},
\end{equation}
and an information rate,
\begin{equation}
R_j(i) = \tau_{ji}  \log \left( 1 + P_j(i) x_{ji}  \right)
= \tau_{ji}  \log \left( 1 + \frac{a_{ji} p_i  \tau_{0i}}{\tau_{ji}}  \right),
\end{equation}
where $a_{ji} = N_0 \, x_{ji}^2$ is an auxiliary fading coefficient\footnote{If the channels are not reciprocal, $a_{ji} = N_0$ $ x_{ji} \, y_{ji}$ where $x_{ji}$ and $y_{ji}$ denote the uplink and downlink channel gains.}.

Since $M\to\infty$, the average WPCN throughout is calculated as $(1/M) \sum_{i=1}^M \sum_{j=1}^K R_j(i)$, and can be maximized according to the following optimization problem \vspace{-1mm}
\begin{eqnarray} \label{rav4}
\underset{p_{i}, \tau_{0i}, \, \tau_{ji}, 1 \leq j \leq K} {\text{max}} \ \frac{1}{M} \sum_{i=1}^M \sum_{j=1}^K \tau_{ji} \log \left( 1 + a_{ji} \, p_{i} \, \frac{\tau_{0i}}{\tau_{ji}} \right) \notag
\end{eqnarray}
\vspace{-5mm}
\begin{eqnarray}
\text{s.t.} &&C1: \frac{1}{M} \sum_{i=1}^M p_i \tau_{0i} \leq P_{avg} \notag \\
&&C2: 0 \leq p_i \leq P_{max}, \forall i \notag \\
&&C3: \sum_{j = 0}^K \tau_{ji} = 1, \forall i \notag \\
&&C4: \tau_{ji} > 0, \forall i j.
\end{eqnarray}


\begin{theorem}
Let us define $a_{ji} = N_0 \, x_{ji}^2$ and $b_i = \sum_{j = 1}^{K} a_{ji}$.

The optimal BS transmit power, $p_i^*$, is determined by
\begin{equation} \label{pi1}
p_i^* =
\begin{cases}
P_{max}, & b_i > \lambda\\
0, & b_i \leq \lambda .
\end{cases}
\end{equation}

The optimal duration of the EH phase, $\tau_{0i}^*$, is found as the root of the following transcendental equation,
\begin{eqnarray} \label{transc}
\log \left(1 + \frac{b_i P_{max} \tau_{0i}^*}{1 -\tau_{0i}^*} \right) + \lambda P_{max} = \frac{b_i P_{max}}{1 - \tau_{0i}^* + b_i P_{max} \tau_{0i}^*} ,
\end{eqnarray}
whereas the optimal IT duration of EHNs $j$ is  given by
\begin{equation} \label{tauij}
\tau_{ji}^* = (1-\tau_{0i}^*) \frac{ \, a_{ji}}{b_i}, \quad 1 \leq j \leq K .
\end{equation}

The constant $\lambda$ is found from $(1/M) \sum_{i=1}^M p_i^* \tau_{0i}^* = P_{avg}$.
\end{theorem}
\begin{IEEEproof}
Please refer to Appendix A.
\end{IEEEproof}

In practice, constant $\lambda$ is estimated by an iterative algorithm, such as the bisection method [\ref{lit12b}]. Basically, the value of $\lambda$ is updated following some rule (e.g., subinterval bisection) until the constraint $C1$ in (\ref{rav4}) is met with some predefined accuracy.

\vspace{-0mm}
\section {Point-to-point EH system with processing cost}
In practical EH transmitters, besides their transmit power, an additional power is also consumed by its non-ideal electric circuitry (e.g., AC/DC converter, analog RF amplifier, and processor), denoted as the {\it processing energy cost} [\ref{lit12a}]. In this section, we extend the system model with non-negligible processing cost to a point-to-point system, consisting of a BS and a single EHN ($K = 1$) and propose a jointly optimal power and time allocation. We consider the following practical model for the total power consumption of the EHN:
\begin{equation}
p_{t} =
\begin{cases} \label{popt}
P_S + p_c, & P_S > 0\\
0, & P_S \leq 0 ,
\end{cases}
\end{equation}
where $P_S$ is the EHN's transmit power, and $p_c$ is the EHN's processing energy cost. Note that the EHN can adapt its transmit power in each epoch, whereas $p_c$ has a fixed value.

The amount of harvested power by the EHN is $E_i = N_0 p_i \, x_i \, \tau_{0i} \, T$, which is completely spent in the successive IT of duration $(1 -  \tau_{0i}) T$ as $E_i = (P_S(i) + p_c) (1 -  \tau_{0i}) T$. Thus, the EHN's transmit power in epoch $i$ is given by
\begin{equation} \label{PS}
P_S(i) = \frac{N_0 p_i x_i \tau_{0i}}{1 - \tau_{0i}} - p_c,
\end{equation}
and the achievable information rate in this epoch is given by  $R_S(i)=(1 - \tau_{0i}) \log \left( 1 + P_S(i) x_i \right)$.

Given $P_{avg}$ and $P_{max}$, we aim at maximizing the average throughput, $(1/M) \sum_{i=1}^M R_S(i)$. In (\ref{rav4}), we set $K = 1$, introduce (\ref{PS}), and restate the optimization problem in terms of the optimization variables $P_S(i)$ and $\tau_{0i}$ as follows:
\begin{equation} \label{PCprob}
\underset{P_S(i), \tau_{0i}, \forall i} {\text{max}} \ \frac{1}{M} \sum_{i=1}^M   \left(1 - \tau_{0i} \right) \log \left( 1 + P_S(i) x_i  \right) \notag
\end{equation}
\vspace{-3mm}
\begin{eqnarray}
\text{s.t.} &C1':& \frac{1}{M} \sum_{i=1}^M \frac{1-\tau_{0i}}{x_i}\left(P_S(i)+p_c \right) \leq  N_0 P_{avg} \notag \\
&C2':& (1-\tau_{0i}) \left(P_S(i)+p_c \right) \leq N_0 P_{max} \tau_{0i} \, x_i, \forall i \notag \\
&C3':& P_S(i) \geq 0, \forall i \notag \\
&C4':& 0 < \tau_{0i} \leq 1, \forall i.
\end{eqnarray}
In (\ref{PCprob}), $C_1'$ and $C_2'$ correspond to $C_1$ and $C_2$ in (\ref{rav4}), $C_3$ and $C_4$ are compressed into the single constraint $C_4'$, whereas $C_3'$ is a natural constraint for EHNs transmit power.

\begin{theorem}
The optimal BS transmit power, $p_i^*$, is determined by
\begin{equation} \label{pi2}
p_i^* =
\begin{cases}
P_{max}, & -\frac{\lambda p_c}{2}+\sqrt{\frac{(\lambda p_c)^2}{4}+\lambda} < x_i < \frac{1}{p_c}\\
0, & \text{otherwise}.
\end{cases}
\end{equation}

The optimal duration of the EH phase, $\tau_{0i}^*$, is given by the root of the following transcendental equation,
\begin{eqnarray} \label{transc2}
 \log \left (1-x_i p_c+\frac{N_0 P_{max} \, x_i^2 \, \tau_{0i}^* }{1-\tau_{0i}^*} \right) + N_0 \lambda P_{max} \notag \\
 = \frac{N_0 P_{max} \, x_i^2}{(1-\tau_{0i}^*) (1-x_i p_c) + N_0 P_{max} \, x_i^2 \, \tau_{0i}^* }.
\end{eqnarray}
\end{theorem}

The constant $\lambda$ is found from $(1/M) \sum_{i=1}^M p_i^* \tau_{0i}^* = P_{avg}$.

\begin{IEEEproof}
Please refer to Appendix B.
\end{IEEEproof}

Note, when setting $p_c = 0$, the optimal solutions (\ref{pi2}) and (\ref{transc2})  reduce to (\ref{pi1}) and (\ref{transc}), respectively, when $K = 1$. This validates our proposed optimal solutions, because (\ref{transc2}) and (\ref{pi2}) are derived based upon an alternative approach to that of the previous section, cf. (\ref{PCprob}) vs. (\ref{rav4}). The respective generalization for the case $K \geq 2$, that involves the processing cost, yields non-convex optimization problem.

\vspace{-3mm}

\section{Numerical results}
We illustrate our results for EH systems with $K = 1$ and $K = 5$ EHNs in Rayleigh fading. The AWGN power is set to $N_0 = 10^{-12}$ W, and the deterministic path loss ($PL$) of the BS-EHN channel is $PL = 60$ dB. Thus, $E[x_{1i}'] = 1/PL$ and $\Omega_1 = 1/(N_0 \, PL) = 10^6$. We consider $M = 10^5$ epochs.

Fig. 2 depicts the average throughput (in bits/symbol) vs. $P_{avg}$ (in watts). Two pairs of curves are presented for: (1) non-zero processing energy cost, $p_c = 10$ $\mu$W (denoted as $p_c > 0$), and (2) zero processing energy cost, $p_c = 0$. The "benchmark" curve refers to the case of a WPCN with fixed BS transmit power ($p_i = P_{avg} = const., \forall i$) and an optimal time allocation according to [\ref{lit8}, Eq. (10)]. Clearly, the joint time and power allocation significantly outperforms the benchmark scheme.

\vspace{-3mm}


\begin{figure}[tbp]
\centering
\includegraphics[width=3.7in]{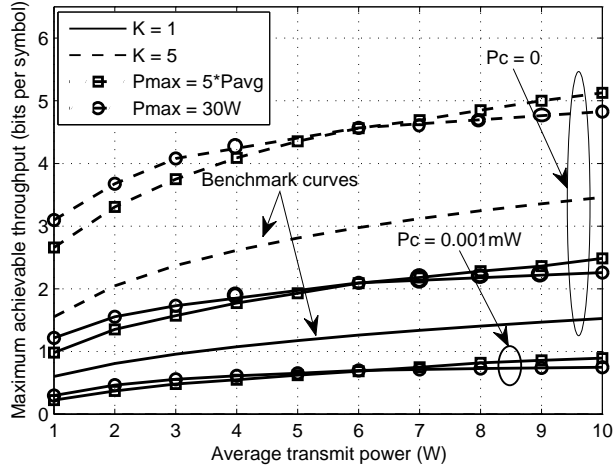} \vspace{-7mm}
\caption{Maximaum achievable throughput vs. BS average output power} \vspace{-3mm}
\label{fig2}
\end{figure}

\appendices

\section{Proof of theorem 1}
The optimization problem (\ref{rav4}) is non-convex because of the products and ratios of the optimization variables $p_i$ and $\tau_{ji}$.  Therefore, we reformulate the problem by introducing the change of variables $e_{i} = p_i \, \tau_{0i}$, and transform (\ref{rav4}) into a convex problem in terms of $e_i$ and $\tau_{ji}$, as
\begin{eqnarray} \label{rav4alt}
\underset{e_i, \tau_{0i} \, \tau_{ji}, 1 \leq j \leq K} {\text{max}} \ \frac{1}{M} \sum_{i=1}^M \sum_{j=1}^K \tau_{ji} \log \left( 1 + a_{ji} \frac{e_{i}}{\tau_{ji}} \right) \notag
\end{eqnarray}
\vspace{-3mm}
\begin{eqnarray}
\text{s.t.} &&\bar C1: \frac{1}{M} \sum_{i=1}^M e_i \leq \bar P \notag \\
&&\bar C2: 0 \leq e_i \leq P_{max} \, \tau_{0i}, \forall i \notag \\
&&C3, \text{ and } C4 \text{ as in } (\ref{rav4}).
\end{eqnarray}
Since $\tau_{ji}$, $1 \leq j \leq K$, do not appear in $C1$ and $C2$ in (\ref{rav4}), (\ref{rav4alt}) can be split into two optimization problems: The first optimizes $\tau_{ji}$, $1 \leq j \leq K$, for given $e_i$ and $\tau_{0i}$ as
\begin{equation}  \label{part1} \vspace{-2mm}
F(e_i, \tau_{0i}) = \underset{\underset{1 \leq j \leq K} {\tau_{ji}>0}} {\text{max}} \ \sum_{j=1}^K \tau_{ji} \log \left( 1 + \frac{a_{ji} \, e_i}{\tau_{ji}}
\right) \notag
\end{equation}
\vspace{-3mm}
\begin{eqnarray}
&& \text{s.t.} \quad \sum_{j=1}^K \tau_{ji} = 1 - \tau_{0i},
\end{eqnarray}
and the second optimizes $e_i$ and $\tau_{0i}$, $\forall i$ given the optimized $\tau_{ji}$, $1 \leq j \leq K$,
\begin{equation} \label{part2} \vspace{-2mm}
\underset{{e_i}, \, \tau_{0i}} {\text{max}} \  \frac{1}{M} \sum_{i=1}^M \, F(e_i, \tau_{0i}),
\end{equation}
\vspace{-3mm}
\begin{eqnarray}
\text{s.t.} &&\bar C1: \frac{1}{M} \sum_{i=1}^M e_i \leq \bar P \notag \\
&&\bar C2: 0 \leq e_i \leq P_{max} \, \tau_{0i}, \forall i \notag.
\end{eqnarray}
Since (\ref{part1}) is convex optimization problem, its Lagrangian is written as
\begin{equation} \label{lagran}
\mathcal{L}_1 = \sum_{j=1}^K \tau_{ji} \log \left( 1 + \frac{a_{ji} \, e_i}{\tau_{ji}} \right) - \mu \left(\sum_{j=1}^K \tau_{ji} + \tau_{0i} - 1 \right),
\end{equation}
where $\mu > 0$ is the Lagrange multiplier associated with the constraint in (\ref{part1}). We now differentiate (\ref{lagran}) with respect to $\tau_{ji}$ and set the derivative to zero, i.e., $\partial \mathcal{L}_1/\partial \tau_{ji} = 0$ for each $j$, $1 \leq j \leq K$, which yields the following set of $K$ equations,
\begin{equation}
\mu = \log \left(1+\frac{a_{ji} e_i}{\tau_{ji}} \right) -
\frac{a_{ji} \frac{e_i}{\tau_{ji}}}{1+a_{ji} \frac{e_i}{\tau_{ji}}}, \,\, 1 \leq j \leq K.
\end{equation}
Given $\mu$, these equations can be satisfied for all $j$, $1 \leq j \leq K$, if
\begin{equation}
\frac{a_{1i}}{\tau_{1i}} = \frac{a_{2i}}{\tau_{2i}} = \frac{a_{3i}}{\tau_{3i}} = ...... = \frac{a_{Ki}}{\tau_{Ki}} = C,
\end{equation}
where $C$ is a function of $\mu$. Considering the constraint in (\ref{part1}), $C$ is found as $C$ $=$ $(1-\tau_{0i})^{-1} \, \sum_{k=1}^K a_{ki}$.

Thus, we obtain the optimal $\tau_{ji}$ is given by (\ref{tauij}). Introducing (\ref{tauij}) into (\ref{part1}), we obtain
\begin{equation} \label{Fopt}
F (e_i, \tau_{0i}) = (1-\tau_{0i}) \log \left(1 + \frac{b_i e_i}{1-\tau_{0i}} \right),
\end{equation}

Now, introducing (\ref{Fopt}) into (\ref{part2}), we obtain a convex optimization problem, whose Lagrangian is given by
\begin{equation*}
\mathcal{L}_2 = \frac{1}{M} \sum_{i = 1}^M (1-\tau_{0i}) \log \left(1+ \frac{b_i e_i}{1-\tau_{0i}} \right) \qquad \qquad \nonumber \\
\end{equation*} \vspace{-3mm}
\begin{equation} \label{lagranprim}
- \lambda \left(\frac{1}{M} \sum_{i = 1}^M e_i - P_{avg}\right) + \sum_{i = 1}^M q_i e_i - \sum_{i = 1}^M  \mu_i (e_i-P_{max} \tau_{0i}).
\end{equation}
where the non-negative Lagrange multipliers $\lambda$, $q_i$ and $\mu_i$ are associated with the constraints $\bar C1$, the left-hand side of $\bar C2$ and the right-hand side of $\bar C2$, respectively.

By differentiating (\ref{lagranprim}) with respect to $\tau_{0i}$ and $e_i$, we obtain:
\begin{eqnarray} \label{partialL1}
\frac{\partial \mathcal{L}_2}{\partial \tau_{0i}}=-\log \left(1+\frac{b_i e_i}{1-\tau_{0i}} \right)+ \nonumber \\
\frac{b_i e_i}{1-\tau_{0i}+b_i e_i} + \mu_i P_{max} = 0
\end{eqnarray}
\begin{eqnarray} \label{partialL2}
\frac{\partial \mathcal{L}_2}{\partial e_i} = \frac{b_i}{1+b_i\frac{e_i}{1-\tau_{0i}}}-\lambda+q_i-\mu_i=0,
\end{eqnarray}
According to the Karush-Juhn-Tucker (KKT) conditions, complementary slackness should be satisfied, $\forall i$: $q_i e_i = \mu_i (e_i-P_{max} \tau_{0i}) = 0$, where $q_i \geq 0$ and $\mu_i \geq 0$. Following a similar mathematical approach as in [\ref{lit12a}, Section III.A], we consider the following 3 cases.

\underline{Case 1}: If $\tau_{0i} = 0$, then $e_i = 0$ and no power is allocated to epoch $i$, i.e., $p_i^* = 0$. Since $e_i=0$, the slackness conditions require $q_i > 0$ and $\mu_i=0$. From (\ref{partialL2}), we obtain the condition $q_i = \lambda - b_i \geq 0$. Thus, this case occurs when  $b_i \leq \lambda$.

\underline{Case 2}: Let us assume $0 < \tau_{0i} < 1$ and  $e_i=P_{max} \tau_{0i}$. This case corresponds to $p_i^* = P_{max}$. The slackness conditions require $q_i=0$ and $\mu_i>0$. From (\ref{partialL2}), we obtain the condition:
\begin{equation} \label{mui}
\mu_i = \frac{b_i}{1 + \frac{b_i P_{max} \tau_{0i}}{1-\tau_{0i}}}-\lambda > 0.
\end{equation}
Introducing (\ref{mui}) and $e_i = P_{max} \tau_{0i}$ into (\ref{partialL1}), we obtain (\ref{transc}). Based upon (\ref{transc}) and (\ref{mui}), it can be shown that the sufficient condition for the occurrence of this case is given by $b_i > \lambda$.

\underline{Case 3}: Let us assume $0 < \tau_{0i} < 1$ and $0 < e_i < P_{max} \tau_i$. Then, the slackness conditions require $\mu_i = q_i=0$. From (\ref{partialL2}), we obtain the equality $e_i = (1-\tau_{0i}) (1/\lambda - 1/b_i)$. Introducing this equality into (\ref{partialL1}) leads to the condition $\lambda/b_i + \log(b_i/\lambda) = 1$, which is satisfied for $b_i = \lambda$.  This case occurs only for one specific value of $b_i$, but its occurrence probability is zero since $x_{ji}$ are continuous random variables. Thus, the optimal $p_i^*$ can be either $0$ (Case 1) or $P_{max}$ (Case 2).

\section{Proof of theorem 2}
The optimization problem (\ref{PCprob}) is non-convex, because the constraints $C_1'$ and $C_2'$ are not convex and the objective function is not concave in $\tau_{0i}$ and $P_S(i)$. Therefore, we reformulate this problem by introducing the change of variables, $\theta_i = 1-\tau_{0i}$ and $\alpha_i  = (1-\tau_{0i}) P_S(i)/x_i$, and transform (\ref{PCprob}) into convex problem is terms of $\theta_i$ and $\alpha_i$,
\begin{equation}
\underset{\alpha_i, \theta_i, \, \forall i} {\text{max}} \ \frac{1}{M} \sum_{i=1}^M   \theta_i \log \left( 1 + \frac{\alpha_i x_i^2}{\theta_i}  \right) \notag
\end{equation}
\vspace{-3mm}
\begin{eqnarray} \label{PCprob2}
\text{s.t.} &\bar C1':& \frac{1}{M} \sum_{i=1}^M \left(\alpha_i + \frac{\theta_i p_c}{x_i} \right) \leq N_0 P_{avg}  \notag \\
&\bar C2':& \alpha_i + \theta_i \left(\frac{p_c}{x_i} + N_0 P_{max} \right) \leq N_0 P_{max}, \forall i \notag \\
&\bar C3':& \alpha_i \geq 0, \forall i \notag \\
&\bar C4':& 0 \leq \theta_i < 1, \forall i.
\end{eqnarray}
The Lagrangian of (\ref{PCprob2}) can be written as
\begin{eqnarray} \label{PClagrange}
\mathcal{L} &=& \frac{1}{M} \sum_{i=1}^M \theta_i \log \left(1+ \frac{\alpha_i x_i^2}{\theta_i} \right) \nonumber \\
&-&\lambda \left(\frac{1}{M} \sum_{i=1}^M \alpha_i + \theta_i \frac{p_c}{x_i} - N_0 P_{avg} \right) \nonumber \\
&-& \sum_{i=1}^M q_i \left(\alpha_i + \theta_i \left(\frac{p_c}{x_i}+N_0 P_{max} \right) - N_0 P_{max}\right)  \nonumber \\
&+& \sum_{i=1}^M \mu_i \alpha_i + \sum_{i=1}^M \upsilon_i \theta_i,
\end{eqnarray}
where the non-negative Lagrangian multipliers $\lambda$, $\mu_i$, $q_i$ and $\upsilon_i$ are respectively associated with $C1'$, $C2'$, $C3'$, and $C4'$ in (\ref{PCprob2}). By differentiating (\ref{PClagrange}) with respect to $\alpha_i$ and $\theta_i$, we get
\begin{equation} \label{PCalpha}
\frac{\partial \mathcal{L} }{\partial \alpha_i}=\frac{x_i^2}{1+\frac{\alpha_i x_i^2}{\theta_i}}-\lambda+\mu_i-q_i=0, \qquad \quad
\end{equation}
\vspace{-3mm}
\begin{eqnarray} \label{PCtheta}
\frac{\partial \mathcal{L} }{\partial \theta_i}=\log \left (1+\frac{\alpha_i x_i^2}{\theta_i}\right)-
\frac{\alpha_i x_i^2}{\theta_i +\alpha_i x_i^2} \qquad \nonumber \\
-\lambda\frac{p_c}{x_i} - q_i \left(\frac{p_c}{x_i}+P_{max}N_0 \right)+\upsilon_i = 0
\end{eqnarray}

According to KKT conditions, the complementary slackness conditions should be satisfied, $\forall i$: $\mu_i \alpha_i = q_i \left[\alpha_i + \theta_i \left(p_c/x_i + N_0 P_{max} \right) - N_0 P_{max} \right] = \upsilon_i \theta_i = 0$, where $\mu_i \geq 0, q_i \geq 0$, and $\upsilon_i \geq 0$. We consider three cases:

\underline{Case 1}: If $\theta_i = 0$, then $\alpha_i = 0$ and no power is allocated to epoch $i$, i.e., $p_i^* = 0$.

\underline{Case 2}: Let us assume $0 < \theta_i < 1$ and $\alpha_i = (1 - \theta_i) N_0 P_{max} - p_c \theta_i/x_i$. This case corresponds to $p_i^* = P_{max}.$ The slackness conditions require $q_i>0$ and $\mu_i = \upsilon_i =0$. From (\ref{PCalpha}), we obtain the condition
\begin{equation} \label{PCq}
q_i=\frac{x_i^2}{1-x_i p_c+\frac{P_{max} (1-\theta_i) N_0 x_i^2}{\theta_i}}-\lambda>0 .
\end{equation}
Introducing (\ref{PCq}) and $\alpha_i = (1 - \theta_i) N_0 P_{max} - p_c \theta_i/x_i$ into (\ref{PCtheta}), we obtain (\ref{transc2}). Based upon (\ref{transc2}) and (\ref{PCq}), it can be shown that the sufficient condition for the occurrence of this case is given by $-\lambda p_c/2 + \sqrt{(\lambda p_c/2)^2 + \lambda} < x_i < 1/p_c.$

\underline{Case 3}: Let us assume $0 < \theta_i < 1$ and $0 < \alpha_i < (1-\theta_i) N_0 P_{max} -p_c \theta_i/x_i$. Then, the slackness conditions require $\mu_i = q_i = \upsilon_i =0$. From (\ref{PCalpha}), we obtain the equality $\alpha_i = \theta_i (1/\lambda - 1/x_i^2)$. Introducing this equality into (\ref{PCtheta}) leads to the equality condition $\lambda (1-p_c x_i)/x_i^2 + \log (x_i^2/\lambda) = 1$. It implies that this case occurs only for one specific value of $x_i$, but its occurrence probability is zero since $x_i$ is a continuous random variable. Thus, the optimal $p_i^*$ can be either $0$ (Case 1) or $P_{max}$ (Case 2).


\begin{thebibliography}{99}

\bibitem{lit1} D. Gunduz, K. Stamatiou, N. Michelusi, and M. Zorzi, "Designing intelligent energy harvesting communication systems," \textit{IEEE Commun. Magazine}, vol. 52, no. 1, pp. 210 - 216, Jan. 2014 \label{lit1}

\bibitem{lit2} C. K. Ho, and R. Zhang, "Optimal energy allocation for wireless communications with energy harvesting constraints," \textit{IEEE Trans. Signal Processing}, vol. 60, no. 9, pp. 4808 - 4818, Sept. 2012 \label{lit2}

\bibitem{lit3} I. Krikidis, T. Charalambous, and J. S. Thompson, "Stability analysis and power optimization for energy harvesting cooperative networks," \textit{IEEE Signal Processing Letters}, vol. 19, no. 1, pp. 20 - 23, Nov. 2011 \label{lit3}

\bibitem{lit4} Z. Ding, and H. V. Poor, "Cooperative energy harvesting networks with spatially random users," \textit{IEEE Signal Processing Letters}, vol. 20, no. 12, pp. 1211 - 1214, Oct. 2013 \label{lit4}

\bibitem{lit6} P. Grover, and A. Sahai, "Shannon meets Tesla: wireless information and power transfer," \textit{Proc. IEEE ISIT 2010}, Austin, TX, USA, June 2010, pp. 2363 - 2367 \label{lit6}

\bibitem{lit7} R. Zhang, and C. K. Ho, "MIMO broadcasting for simultaneous wireless information and power transfer," \textit{IEEE Trans. Wireless Commun.}, vol. 12, no. 5, pp. 1989 - 2001, May 2013 \label{lit7}

\newpage

\bibitem{lit8} H. Yu, and R. Zhang, "Throughput maximization in wireless powered communication networks," \textit{IEEE Trans. Wireless Commun.}, vol 13, no. 1, pp. 418-428, Jan 2014 \label{lit8}

\bibitem{lit9} H. Ju, and R. Zhang, "Optimal resource allocation in full-duplex wireless-powered communication network," \textit{IEEE Trans. Commun.}, vol. 62, no. 10, pp. 3528 - 3540, Oct. 2014 \label{lit9}

\bibitem{lit10} X. Kang, C. Keong Ho, and S. Sun, "Optimal time allocation for dynamic-TDMA-based wireless powered communication networks," \textit{Proc. IEEE Globecom 2014, Signal Processing for Communications Symposium}, Austin, TX, USA, Dec. 2014 \label{lit10}

\bibitem{lit11} R. Morsi, D. S. Michalopoulos, and R. Schober, "Performance analysis of wireless powered communication with finite/infinite energy storage," arXiv:1410.1805v2, Oct 2014. \label{lit11}

\bibitem{lit12} He Chen, et al., "Harvest-then-cooperate: wireless-powered cooperative communications", \textit{IEEE Trans. Signal Processing}, vol. 63, no. 7, pp. 1700 - 1711, Apr. 2015 \label{lit12}

\bibitem{lit12a} O. Orhan, D. Gunduz, E. Erkip, "Throughput maximization for an energy harvesting communication system with processing cost", \textit{Proc. 2012 IEEE Information Theory Workshop}, Lausanne, Sept. 2012, pp. 84-88 \label{lit12a}

\bibitem{lit12b} S. Boyd, EE364b Lecture Notes. Stanford, CA: Stanford Univ. Available: http://stanford.edu/class/ee364b/lectures/ellipsoid\_method\_notes.pdf \label{lit12b}

\end{thebibliography}
\end{document}